RESEARCH ARTICLE                                                                                                    OPEN ACCESS

# Analysis of Attacks on Hybrid DWT-DCT Algorithm for Digital Image Watermarking With MATLAB

Lalit Kumar Saini[1], Vishal Shrivastava[2]
M.Tech Scholar[1], Professor[2]
Department of Computer Science and Engineering
Arya College of Engineering and Information Technology,
Jaipur, India

**ABSTRACT**
Watermarking algorithms needs properties of robustness and perceptibility. But these properties are affected by different -2 types of attacks performed on watermarked images. The goal of performing attacks is destroy the information of watermark hidden in the watermarked image. So every Algorithms should be previously tested by developers so that it would not affected by attacks.

## I. INTRODUCTION

Digital image watermarking is process of adding some information in image in form of text, image and logo for the purpose of owner identification and security. That information may be in visible or in invisible form. So that anyone can be extracted that info when required for a particular purposes. The main obstacle in extraction process of watermark information is the different-2 attacks perform of watermarked image. These attacks degrade the watermark information embedded in watermarked image. Sometimes they affected so much as the watermark information will be destroyed. This paper focuses on how to analyse the watermarked image with MATLAB and the effect of attacks on a DWT-DCT hybrid algorithm. For this a GUI tool is designed in MATLB2012b for convenience.

The paper is organized as follows sections:

- Overview the Hybrid DWT-DWT algorithm a
- Results of Hybrid Algorithm
- Overview and Implementation of Attacks
- Analysis and Results of attacks on Algorithm

## II. OVERVIEW OF GENERAL & HYBRID DWT- DCT HYBRID ALGORITHM

In an Image watermarking anyone can embedded information in form of Visible or in Invisible form. There are so many types of watermarking methods.

*2.1 General Watermarking Techniques*

Some of the general watermarking techniques are as follows

*2.1.1 Spatial domain watermarking*: In this watermarking the watermarking information is hidden in the least significant bit (LSB) of cover image. Due to Changes in LSB the effect on cover image is very little, so that anyone not recognized it generally.

*2.1.2 Transfer Domain Watermarking*: In this watermarking method the watermarking information is hidden in the coefficients of transfer domain. These coefficients may be from the wavelet transfer domain, Frequency transfer domain, Cosine transfer domain.

*2.1.3 Hybrid domain watermarking*: In this type of watermarking anyone can combined any of the above techniques to form a new technique. The advantage of hybrid method is disadvantages of one method are removed when combined with others. Some of the most used combinations are:

A. DCT-DWT-SVD
B. DWT-DCT
C. SVD-DCT
D. SVD-DWT

Some of researches done great jobs over last few years on the combinations of above these and find the suitable results.
Some of the Combinations work as follows:

Saeed K. Amirgholipour, Ahmad R. Naghsh-Nilchi [1] combines DWT-DCT transforms for performing watermarking.

S Mukherjee, Arup K. Pal [2] combines SVD-DCT

transformations to get better results.

Parthiban V, Ganesan R [3] combines the DWT-SVD transform to embedding watermarking.

Md. Maklachur Rahman [4] combines all three techniques as SVD-DCT-DWT and found the better results.

Lalit k. Saini,Vishal Shrivastava[5] also combines the DWT-DCT transfer domain to get better results for these combination.

*2.2 Hybrid DWT-DCT watermarking method*

As reference a Hybrid DWT-DCT [5] method of watermarking is taken.
In this method 512*512 dimensional gray scale images is used as cover image and 32*32 dimensional image is used as watermark logo. Initially cover image is DWT transform up-to 3 levels to get four HH sub-bands. These sub-bands then convert into 4*4 blocks to get DCT bands. Then binary converted watermark logo is embedded in the cover image with embedded algorithm. All the process is reversed to get watermarked image.
Same procedure DWT then DCT is doing in extraction algorithm for extraction of watermark logo.

## III. RESULTS OF HYBRID DWT-DCT IMAGE WATERMARKING ALGORITHM

The results of above hybrid algorithm are impressive as getting PSNR value more than 40db which is acceptable value for any good watermarking algorithm. So that the algorithm is perceptible and robust against visual attacks.

Results for image leena.bmp is

TABLE 1: Results for Cover Image Leena.bmp

| S.no | Cover Image | Extracted Watermark | MAE | PSNR |
|---|---|---|---|---|
| 1 | 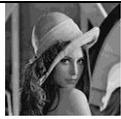 | 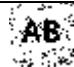 | 1.0681e-04 | 44.0007 |

**Overview and Implementation of Attacks With MATLAB**

*3.1 Overview of Attacks on Watermarking*

Attacks are the factors or processes that can degrade the digital watermark strength. Attacks can be broadly classified into these main categories.

- Removal Attacks
- Geometric Attacks
- Cryptographic Attacks
- Protocol Attacks.

*3.1.1 Removal Attacks*

These type of attacks are affected the watermark in such a way that it a complete or nearly about to removed or destroyed watermark data. Example of these attacks are denoising, quantization (e.g., for compression), remodulating, and collusion attacks.

*3.1.2 Geometric Attacks*

These types of attacks can attacks the pixels of image for attacking. Like pixels shifting, scaling of image, rotation of image without any higher visual changes. The aim of these kinds of attacks is degrade the quality of watermark.

*3.1.3 Cryptographic Attacks*

In these kinds of attacks the attackers are finding the loopholes in main embedding algorithm and remove the watermark information. Examples are brute force attack and oracle attack.
But if the embedding algorithm is complex then these attacks are easily restricted.

*3.1.4 Protocol Attacks*

These attacks are intentionally done by attackers to change or destroy the ownership information from the watermarked image. Example of these attacks is copy attack and changing of watermark.

Beyond these attacks there are so many new attacks continuously developed by hackers to affect the watermarking algorithms and watermark.

*3.2 Implementation of some attacks with MATLAB*

MATLAB is a widely used as a research and simulation tool in universities, research labs and industry. MATLAB has great graphical properties both in 2D or 3D graphical research. This section describes how to implementing and analysis the various type of attacks [6] and their effect of watermarking scheme.

*3.2.1 Gaussian noise Attack*
Gaussian noise is also referred as white noise. This can be implementing in MATLAB with function

AI = mnoise( image,'gaussian',mean_value, variance)

Where AI= Image after attack
image = Source Image
gaussian = Keyword for perform Gaussian noise attack
mean_value = vaule of a mean for Gaussian noise attack
variance = Value of a variance for Gaussian noise attack

3.2.2 Peeper salt noise attack
It's a type of attack in which balk and white pixels present in the image as a noise. The effect of this attack can be seen on image by following MATLAB function

   AI = imnoise(image,'salt & pepper',den);

Where AI= Image after attack
image = Source Image
salt & pepper = Keyword for perform salt & pepper noise attack
den = value of noise density for attack

### 3.2.3 Speckle noise Attack
Speckle noise [7] is commonly found in satellite images, medical images and synthetic aperture radar images. It is a Multiplicative noise. This can be implementing in MATLAB with function

   AI = mnoise( image,'speckle',variance)

Where AI= Image after attack
image = Source Image
speckle = Keyword for perform speckle noise attack
variance = Value of a variance for speckle noise attack

3.2.4 Intensity Transformation
It's a type of attack in which attacker change the intensity of the watermarked image to degrade the watermark information. its types and implementation with MATLAB are given below.

- photographic negative (using imcomplement function)
- gamma transformation (using imadjust)
- logarithmic transformations (using c*log(1+f))
- contrast-stretching transformations (using 1./(1+(m./(double(f)+eps)).^E)

Gamma intensity transformation function

AI= imjust (image, [], [], gamma value)

Where AI= Image after attack
Gamma Value= value of gamma for attack if <1 brighten the intensity and if >1darken the intensity of grayscale component.

### 3.2.5 Burred image attack
Image is burred due to many factors like low focal quality of camera, atmospheric effects etc. Hackers use it generally for attacking watermarked image to degrade the quality of watermark.
In MATLAB there is function or command **fspecial** for that. The function imfilter can then be used to blur the image.

   H = fspecial('gaussian', size(image), SD);
   IA= imfilter(image, H);

   AI = imnoise (image, 'salt & pepper', density);

Where AI= Image after attack
image = Source Image
SD = Standard deviation

### 3.2.6 Image contrast attack
Changing in contrast [8] of image would affect the image and watermark also.
This can be implementing in MATLAB with function

   AI = histeq (image);

or by using image contrast adjustment toll opened in separate window by
   Imcontrast(h)

imcontrast( h) creates the Adjust Contrast tool associated with the image specified by the handle h.

Where AI= Image after attack
image = Source Image

Beyond all above there are so many types of attacks like resizing, cropping, scaling, sharpening, JPEG compression etc. which affects the quality of watermark image and watermark too. The effect of these all attacks are easily analyzed with the help of MATLAB functions specially for this purpose. For analyze the algorithm firstly attack the image with any of these attack. After that recover the watermark information from attacked image. Compare the quality of logo recovered from non-attacked and recovered from attacked image. Thus anyone can analyze the strength of algorithm against these attacks.

## IV. RESULTS OF ATTACK ON HYBRID DWT-DCT ALGORITHM

To analyze the different-2 attacks on the hybrid DWT-DCT [5] algorithm anyone can recover the watermark logo of algorithm with same source image. But the condition is at first attack the watermarked image with any attack and then recovered the logo with hybrid extraction algorithm. Then analyze the changes in the quality of watermark logo.
To evaluate the capability of algorithm against different-2 attacks a GUI tool designed in MATLAB2012b for convenience shown in figure 1.

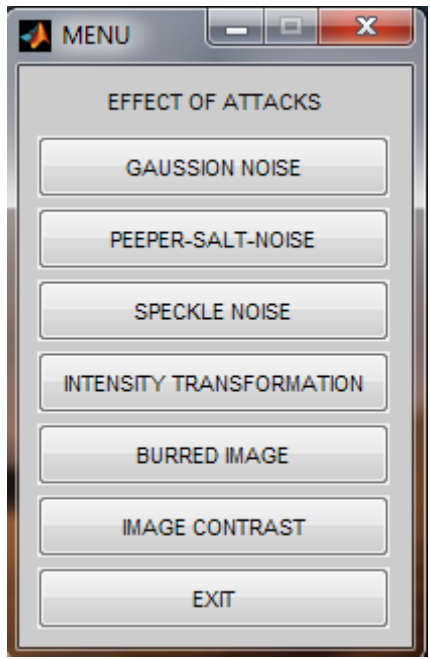

Fig 1 GUI tool for analyze of attacks on recovered watermark

In this tool anyone can press the particular button for attacking watermarked image and recovered watermark from attacked image.

The results of effect of attacks on recovered logo after attacking watermarked image shown in table 2.

TABLE 2: Effect of different-2 attacks on extracted watermark

| S.no | Type of Attack | Effect on Extracted Watermark logo |
|---|---|---|
| 1. | Gaussion Noice | |
| 2. | Peeper & Salt Noice | |
| 3. | Speckle Noice | |
| 4. | Intensity Transformation | |
| 5. | Burred Image | |
| 6. | Contrast Change | |

As shown in the table 2 that all the above attacks affects the quality of recovered watermark logo. But not in such a way that it cannot be recognized.

## V. CONCLUSION

All the attacks are the threats for every watermarking algorithms and techniques. They degrade or destroy the watermark information. So the watermarking algorithm should be as robust as it can be resist against these attacks. MATLAB is scientific tool by which researchers can easily tests the strength of any particular algorithm against these attacks. The described Hybrid DWT-DCT algorithm [5] had a capability of resistance against these attacks.

## REFRENCES

[1.] Saeed K. Amirgholipour, Ahmad R. Naghsh-Nilchi, "Robust Digital Image Watermarking Based on Joint DWT-DCT", International Journal of Digital Content Technology and its Applications Volume 3, Number 2, June 2009

[2.] S. Mukerjee, Arup K. Pal ," A DCT-SVD based robust watermarking scheme for grayscale image", Proceeding, ICACCI12,ACM , NY,USA

[3.] Parthiban V, Ganesan R, " Hybrid Watermarking Scheme for Digital Images"Journal of Computer Applications ISSN: 0974 – 1925,Volume-5, Issue EICA2012-1, February 10, 2012

[4.] Md. Maklachur Rahman, "A DWT,DCT and SVD based watermarking technique to protect the image piracy", International Journal of Managing Public Sector Information and Communication Technologies (IJMPICT)Vol. 4, No. 2, June 2013

[5.] Lalit Kumar Saini, Vishal Shrivastava, "A New Hybrid DWT-DCT Algorithm for Digital Image Watermarking", International Journal of Advance Engineering and Research Development (IJAERD) , Volume 1,Issue 5,May 2014

[6.] Baisa L Gunjal, Dr. Suresh N Mali, "Handling Various Attacks in Image Watermarking", CSI Communications , 2013

[7.] Milindkumar V. Sarode , Prashant R. Deshmukh "Reduction of Speckle Noise and Image Enhancement of Images Using Filtering Technique", International Journal of Advancements in Technology, Vol 2, No 1 (January 2011),pp.30-38

[8.] http://www.mathworks.in/help/images/ref/imcontrast.html